\documentclass[9pt,a4paper,twocolumn]{article}

\usepackage[utf8]{inputenc}
\usepackage{dcolumn}% Align table columns on decimal point
\usepackage{bm}% bold math
\usepackage{braket}
\usepackage[T1]{fontenc} 
\usepackage{float}
\usepackage{amsmath,amsfonts,amssymb,eqnarray}  
\usepackage{color,soul}
\usepackage{graphicx}  
\usepackage{array,booktabs}
\usepackage[pdftex,hypertexnames=false]{hyperref} 
\usepackage[affil-it]{authblk}
\usepackage[font=small,labelfont=bf]{caption}
\usepackage{cite}
\usepackage[left=2cm,right=2cm,top=1.6cm]{geometry}
\usepackage{titlesec}
\usepackage{multirow}
\definecolor{mycol}{RGB}{255,230,204}
\usepackage{mdframed}

\titleformat{\section}{\normalfont\fontsize{13}{15}\bfseries}{\thesection}{0.5em}{}

\title{\textbf{Path-encoded high-dimensional quantum communication over a 2 km multicore fiber}
} 

\author{Beatrice Da Lio$^{1}$, Daniele Cozzolino$^{1}$, Nicola Biagi$^{2}$, Yunhong Ding$^{1}$, Karsten Rottwitt$^{1}$, Alessandro Zavatta$^{2,3}$, Davide Bacco$^{1*}$, Leif K. Oxenl\o we$^{1}$}

\affil{\small{
$^\textrm{\textit{1}}$Center for Silicon Photonics for Optical Communication (SPOC), Department of Photonics Engineering, Technical University of Denmark, 2800 Kgs. Lyngby, Denmark.\\
$^\textrm{\textit{2}}$CNR - Istituto Nazionale di Ottica (CNR-INO), Largo E. Fermi, 6 - 50125 Firenze, Italy.\\
$^\textrm{\textit{3}}$LENS and Dipartimento di Fisica e Astronomia, Università di Firenze, Via G. Sansone, 1 - 50019 Sesto Fiorentino, Italy.\\* dabac@fotonik.dtu.dk
}}
\date{}

\begin{document}
\maketitle

\begin{abstract}
Quantum key distribution (QKD) protocols based on high-dimensional quantum states have shown the route to increase the key rate generation while benefiting of enhanced error tolerance, thus overcoming the limitations of two-dimensional QKD protocols. Nonetheless, the reliable transmission through fiber links of high-dimensional quantum states remains an open challenge that must be addressed to boost their application.
Here, we demonstrate the reliable transmission over a 2 km long multicore fiber of path-encoded high-dimensional quantum states. Leveraging on a phase-locked loop system, a stable interferometric detection is guaranteed, allowing for low error rates and the generation of 6.3 Mbit/s of secret key rate.
\end{abstract}

\section{Introduction}
Quantum key distribution (QKD) constitutes the very first step towards a quantum internet and it is the most technologically advanced application in quantum communication so far~\cite{wehner2018quantum,pirandola2019advances}. It allows two remote users to exchange secret keys, used to encrypt and decrypt their data communications, in an information theoretic secure way thanks to the laws of quantum physics.
State of the art experiments directly address the factors that are currently limiting the actual deployment of QKD technology. Mainly, these factors are the achievable communication distance~\cite{chen2020sending,boaron2018secure,yin2016measurement}, the key rate generation~\cite{islam2017provably,bacco2019boosting} and the coexistence of QKD protocols with classical communication channels~\cite{bacco2019boosting,wang2020long}.
A way to face the key rate generation problem is to perform high-dimensional QKD protocols. Indeed, a quantum state spanning a $d$-dimensional Hilbert space, a \textit{qudit}, being able to encode $\log_2(d)$ classical information bits, owns a larger information capacity with respect to a qubit.
Moreover, it has been shown that high-dimensional states possess a higher noise resilience, which implies a higher error tolerance in a QKD session if compared to the qubit case~\cite{cozzolino2019high,sheridan2010security,ecker2019overcoming}. Different photonic degrees of freedom can be exploited to prepare high-dimensional states, such as the orbital angular momentum of light~\cite{cozzolinopra2019,giordani2019experimental,cozzolino2019air,dixon2012quantum}, frequency~\cite{Kues2017,Jin_2016}, time-energy and time-bin encoding~\cite{islam2017provably,mower2013high,ali2007large,bunandar2015practical,Steinlechner2017,Martin2017} and path~\cite{Wang2018,Krennpath2017,Ding2017,adcock2019programmable,llewellyn2020chip}. Each degree of freedom offers different advantages in terms of stability, control and scalability, while facing different problems~\cite{cozzolino2019high}. Indeed, the orbital angular momentum of light can be considered as a natural choice to enlarge the Hilbert space, however, the on-chip integration of its generation and manipulation devices is very demanding and makes its scalability low. Time-energy and time-bin encoding are perhaps the simplest and most used approaches for generating and distributing high-dimensional quantum states. Nonetheless, increasing dimensions by using time will limit the overall performance of the quantum systems since the repetition rate of the generated states would rapidly decrease, and this could be a non‐trivial issue for technological applications. Path-encoding is a very promising approach due to its very good compatibility with photonic integrated circuits. Indeed, the ease to generate, manipulate, and detect path-encoded quantum states has made path-encoding largely used to achieve groundbreaking results~\cite{Wang2018,Ding2017,luo2019chip,llewellyn2020chip}.
However, the major challenge of this approach is the reliable transmission of such states. A first solution could be to couple each path to a single mode fiber (SMF), but each of them would experience different random phase drifts due to temperature changes, bends, and mechanical stress, increasingly disrupting the transmission of the superposition states with longer communication channels. Indeed, the transmission of these states implies stable and well defined phase relations to be maintained to accomplish the interferometric measurement. An alternative approach is to couple each path to a different core of a multicore fiber (MCF). Since all the cores are enclosed in the same cladding area, phase drifts among them are highly suppressed, thus allowing for a better transmission, especially of the superposition states~\cite{xavier2020quantum,dalio2019}. Nonetheless, accounting for the different phase drifts still remains vital for the stability of the communication system. In fact, previous experiments already investigated these fibers as a mean for high-dimensional quantum communication, but their limitations in terms of stability affected the achievable distance~\cite{Ding2017,canas2017,lee2017experimental}. Recently, we have approached this matter by implementing an active stabilization of the system~\cite{dalio2019}. In particular, we have demonstrated the high fidelity transmission of \textit{ququarts} (quantum states with $d=4$) over a 2 km long MCF, extending the reach of previous works~\cite{Ding2017, canas2017}.\\
In this work, we extend our previous results by implementing a scalable scheme allowing for real-time state modulation needed to perform a high-dimensional high-speed QKD protocol. Our solution shows high reliability and long term stability, as it maintains a persistent phase difference over 2 km long fiber interference for several hours of continuous and free running acquisition. These characteristics allow for a secret key generation rate of 6.3 Mbit/s.

\section{Results}
\subsection{Phase modulation and stabilization}
%Recap phase drifts
As outlined in the introduction, the faithful transmission of path-encoded qudits is affected by random phase drifts among the different paths, disrupting the phase coherence of superposition states. Exploiting the slower phase drift rate experienced through the cores of a MCF than those between different SMFs~\cite{dalio2019}, MCFs are a more fitting and an easier to deal with platform to transmit path-encoded quantum states. Nevertheless, the presence of phase drifts requires the adoption of a stabilization system to compensate them, thus ideally eliminating this source of error intrinsic to the channel. Two distinct implementations have been approached so far: in the first, the stabilization loop utilizes the same error on the quantum states as a reference signal to drive an actuator to compensate for such errors~\cite{canas2017}, meaning that a stabilization routine must be carried out switching off the QKD session momentarily. In the second method, the stabilization channel and the quantum states transmission are independent and can run simultaneously, which require a proper multiplexing of the two signals~\cite{dalio2019}. 
In particular, in our previous work we reported the distribution of single quantum states over 2 km of multicore fiber, using a counter-propagating signal together with a phase-locked loop (PLL) board to stabilize the channel phase drifts. More details on the PLL board are reported in the Supplementary Information. A possible drawback of this method is the impossibility of correcting faster phase drifts that can arise in long-haul links:
\begin{figure}[ht!]
\centering
\includegraphics[width=0.47\textwidth]{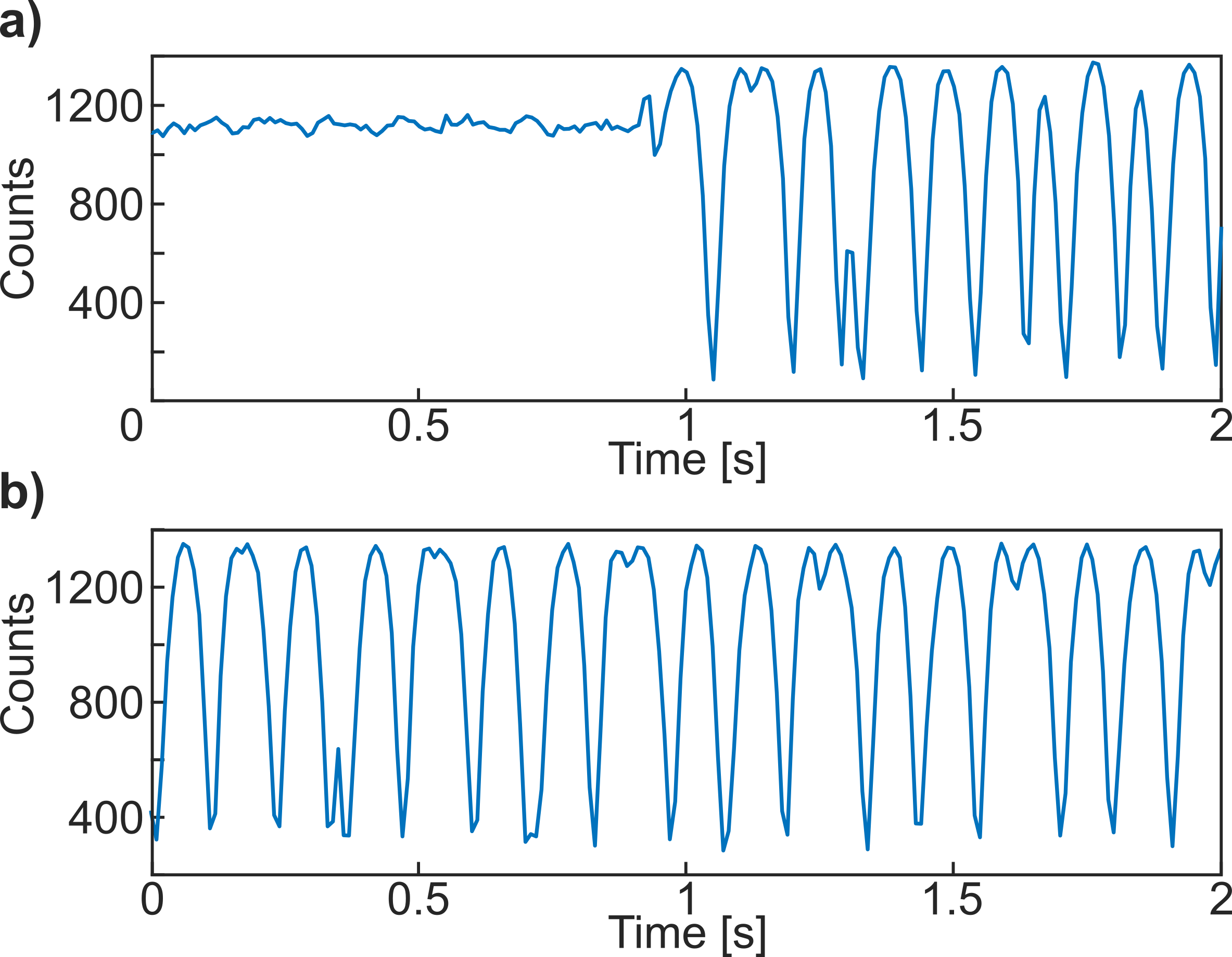}
\caption{\textbf{a) Polarization aligned to the phase modulator.} Acquired counts from the stabilization channel when the phase-locked loop board is inducing interference fringes, and the stabilization signal is aligned with the modulated polarization in the phase modulator. The phase modulator is initially on and it is turned off after approximately one second: the interference fringes are not visible when the stabilization is phase modulated. \textbf{b) Polarization orthogonal to the phase modulator.} Acquired counts from the stabilization channel with orthogonal polarization and phase modulator turned on: the interference fringes induced by the board are visible, as the signal is not phase modulated.}
\label{fig:fringes}
\end{figure}
due to the finite speed of light, the two counter-propagating signals can be affected by different accumulated phase drifts. Here, we present a stabilization system based on two different co-propagating signals that can overcome this issue, benefiting from a simpler and more scalable implementation as well. Moreover, we integrate the stabilization channel with a fast optical phase modulation of the quantum states required for random state choice.
%Modulation real-time + reason loop
Indeed, the high-dimensional QKD protocol that is realized in this work requires the preparation of the following two mutually unbiased bases:
\begin{equation}
    \mathcal{Z} = \frac{1}{\sqrt{2}} \begin{pmatrix} \ket{1}+\ket{5}\\
    \ket{1}-\ket{5}\\
    \ket{7}+\ket{2}\\
    \ket{7}-\ket{2}
    \end{pmatrix}
    \:
    \mathcal{X} = \frac{1}{\sqrt{2}} \begin{pmatrix} \ket{1}+\ket{7}\\
    \ket{1}-\ket{7}\\
    \ket{5}+\ket{2}\\
    \ket{5}-\ket{2}
    \end{pmatrix}\;,
    \label{eq_states}
\end{equation}
where a state $\ket{k}$ is encoded in the $k$-th core of the MCF. As the MCF used in this experiment has seven cores, we chose to use the four cores with lower loss and cross-talk~\cite{dalio2019}.
\begin{figure*}[ht!]
\centering
\includegraphics[width=1\textwidth]{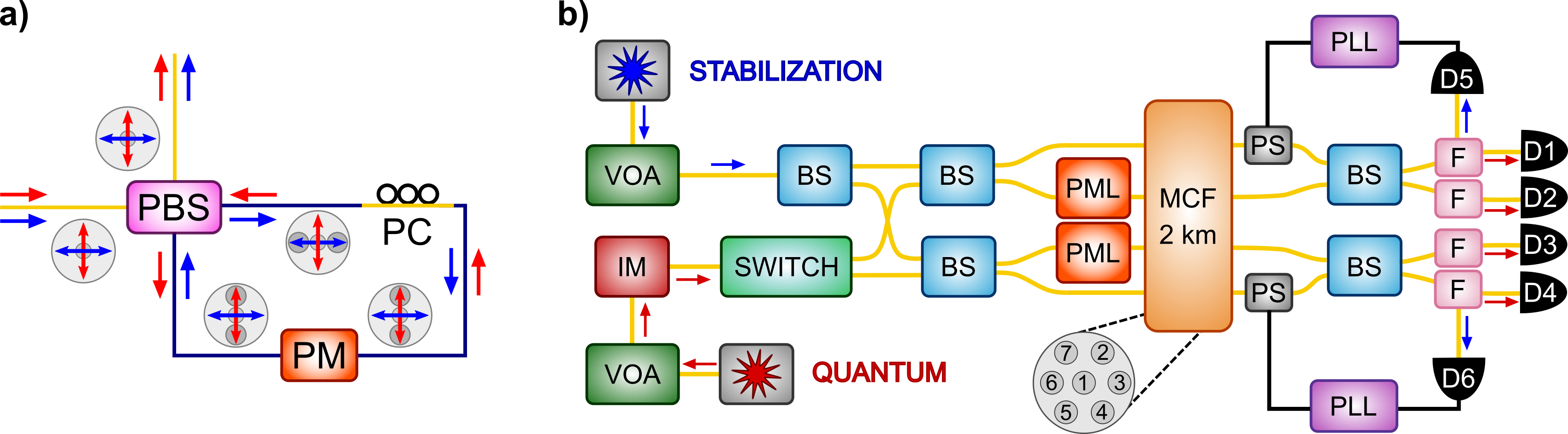}
\caption{\textbf{a) Phase modulation loop.} Red arrow: quantum channel; blue arrow: stabilization channel; yellow fiber: single mode fiber; blue fiber: polarization maintaining fiber; PBS: polarizing beam splitter; PM: phase modulator; PC: polarization controller.
\textbf{b) Setup scheme.} Red arrow: quantum channel; blue arrow: stabilization channel; VOA: variable optical attenuator; IM: intensity modulator; BS: beam splitter; SWITCH: optical switch; PML: phase modulation loop, see figure~\ref{fig:setup} a); MCF: multicore fiber; PS: phase shifter; PLL: phase-locked loop board; F: wavelength division multiplexing filters; D1, D2, D3 and D4: superconducting nanowire single photon detectors; D5 and D6: InGaAs single photon detectors.}
\label{fig:setup}
\end{figure*}
All eight states live on a superposition of two cores with $0$ or $\pi$ phase difference: this particular choice allows to simplify both the quantum and the stabilization systems actual implementations, as demonstrated in our previous work~\cite{dalio2019}. Such phase relations can be experimentally implemented with fast optical phase modulators. However, the integration of these devices is a non-trivial issue: to effectively stabilize phase drifts, the stabilization channel needs to be transmitted in the same fiber interferometer the quantum states propagate through. In other words, both signals need to travel along the same optical path. Hence, the stabilization channel would acquire the same fast phase modulation of the quantum states. Such modulation hinders the ability of the system to track the random phase drifts, as they are several orders of magnitude slower than the modulation. 
%Loop + Figure 1a
\noindent
This effect is shown in figure \ref{fig:fringes} a), where the phase modulator is initially on and it is turned off after approximately one second. As it can be seen, the interference fringes induced by the PLL board are not visible when the stabilisation signal is phase modulated. To solve this issue and be able to effectively phase modulate only the quantum channel while both signals propagate through the same fiber paths, we exploit the polarization dependence of phase modulator crystals. Indeed, by orienting the polarization of the stabilization signal orthogonally to the modulation axis of the phase modulator, the output is poorly affected by the modulation. This is shown in figure \ref{fig:fringes} b), where the PLL board induced fringes are visible despite the action of the phase modulator. Hence, we designed the phase modulation loop (PML) shown in figure~\ref{fig:setup} a). The quantum channel, represented with a red arrow, is vertically polarized at the input of a polarizing beam splitter (PBS), whereas the stabilization channel, indicated with a blue arrow, is horizontally polarized. The PBS splits the two signals: the first is reflected while the second transmitted. By connecting the two outputs of the PBS, we obtain a loop in which the quantum signal travels in a counter-clockwise direction while the stabilization signal in a clockwise direction. To be noted that we use a fiber-PBS: by design, at both outputs the signals will be aligned to the slow axis of the polarization maintaining fibers. As the phase modulator (PM) efficiently modulates only the mode aligned with the slow axis of the polarization maintaining fiber, we place it at the reflected output of the PBS. This ensures the correct modulation of the quantum channel. Contrariwise, on the other output of the PBS, the stabilization channel (also aligned to the slow axis) needs to be rotated to the orthogonal fast axis before it can be sent through the PM to avoid modulation. This rotation can be achieved by inserting in the loop a polarization controller (PC). Hence, when the two channels reach the PBS after one loop, they are both directed to the secondary input of the PBS.\\
%Setup + phase-locked loop + Figure 1b
The PML just described is integrated in the setup as part of the quantum state preparation, as shown in figure~\ref{fig:setup} b). The transmitter, called Alice, has to prepare the quantum states belonging to the two bases in eq.~\eqref{eq_states} and send them, together with a stabilization signal, through the MCF towards the receiver, Bob. Bob's tasks are to both measure the states via projective measurements, and actively stabilize the transmission channel. An in-depth description of the experimental setup is reported in the Methods section.

\begin{figure*}[ht!]
\centering
\includegraphics[width=1\textwidth]{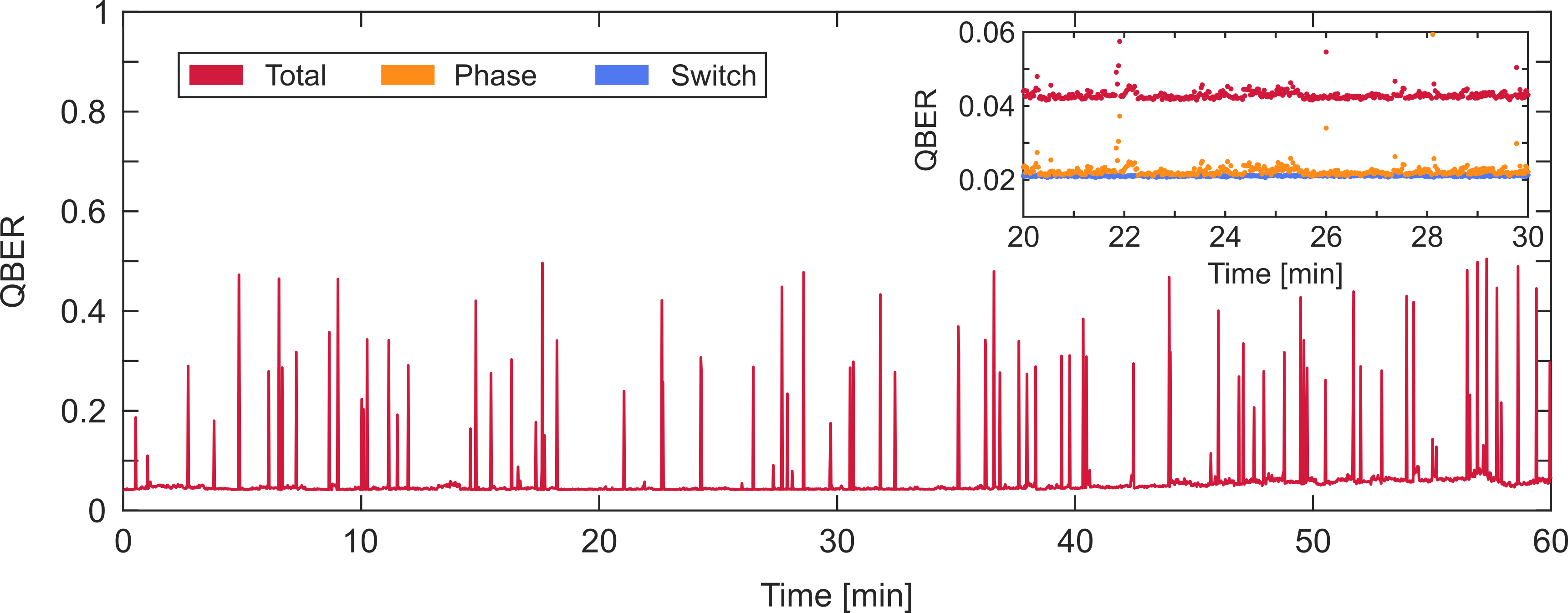}
\caption{\textbf{System stability.} Measured QBER in the $\mathcal{X}$ basis over one hour of continuous acquisition. Vertical lines show moments when the tracking system lost its locking position, yet recovering right away the previous stable QBER value. The inset show a magnification of the same acquisition from minute 20 to 30: in red the same measured QBER, with average value 4.9\%, and in orange and blue its two contributions due to errors in the phase modulation and stabilization (average 2.8\%) and to the switch modulation (average 2.1\%) respectively.}
\label{fig:stab}
\end{figure*}

\subsection{Quantum protocol}
%Stability + QBER in time
We show the typical behaviour of the setup in figure~\ref{fig:stab}, where one hour of free-running continuous data acquisition is reported. The system behaviour over more than 7 hours is reported in the Supplementary Information. In figure~\ref{fig:stab}, the QBER shown in red is measured with the system in the $\mathcal{X}$ basis configuration, with an average photon number per pulse $\nu \approx 0.24$ photon/pulse. The measurement demonstrates the ability of the stabilization system to track and compensate for the random phase drifts accumulated during the 2 km fiber transmission, and hence to successfully maintain a stable QBER value during the whole acquisition, with an average of 4.9\%. The abrupt changes, appearing as vertical lines, denote the moments where at least one of the PLL boards lost its locking position. Nevertheless, the system is able to recover the previous position with similar QBER performance almost instantaneously. The slow increase in QBER happening after approximately 40 minutes is most likely due to temperature changes in the lab and/or polarization drifts.
\begin{table*}[ht!]
\begin{mdframed}[backgroundcolor=mycol,innerleftmargin=0pt,topline=false,rightline=false,leftline=false,bottomline=false]
\caption{Parameters experimentally used, measured QBER in all four possible configurations of basis and intensity choice, and obtained secret key rate values.}
    \label{tab:par_result}
    \centering
    \begin{tabular}{p{4 cm} p{1.4cm} p{1.4cm} p{1.4cm} p{1.4cm} p{1.4cm} p{1.4cm}}
    \toprule
         \textbf{Channel loss [dB]} & 5.8 & 9.8 & 13.8 & 17.8 & 21.8 & 25.8 \\[1ex]
           \hline\addlinespace[1ex]
         $\mu_1$ [photon/pulse] & 0.19 & 0.20 & 0.22 & 0.23 & 0.23 & 0.22 \\[1ex]
         $\mu_2$ [photon/pulse] & 0.15 & 0.16 & 0.17 & 0.18 & 0.18 & 0.18 \\[1ex]
         p$_{\mu_1}$ & 0.62 & 0.63 & 0.63 & 0.63 & 0.63 & 0.64 \\[1ex]
         p$_{\mathcal{Z}}$ & 0.90 & 0.90 & 0.90 & 0.90 & 0.90 & 0.86 \\[1ex]
         \hline\addlinespace[1ex]
         QBER$_{Z\mu_1}$ & 4.32\% & 4.66\% & 4.67\% & 5.10\% & 5.84\% & 6.98\% \\[1ex]
         QBER$_{Z\mu_2}$ & 4.10\% & 4.81\% & 4.62\% & 5.08\% & 5.72\% & 7.58\% \\[1ex]
         QBER$_{X\mu_1}$ & 4.73\% & 4.46\% & 4.99\% & 5.09\% & 5.94\% & 7.48\% \\[1ex]
         QBER$_{X\mu_2}$ & 4.66\% & 4.83\% & 4.99\% & 5.16\% & 6.28\% & 8.28\% \\[1ex]
         R$_{sk}$ [kbit/s]& 6308 & 2585 & 796 & 258 & 116 & 22 \\
         \bottomrule
    \end{tabular}
    \end{mdframed}
\end{table*}
The inset in figure~\ref{fig:stab} shows the same QBER acquisition from minute 20 to 30, also presenting the typical contributions to the overall measured QBER value. Indeed, apart from the random phase drifts experienced during fiber transmission, the optical switch and the phase modulators constitute sources of errors. The measured QBER due to only the switch modulation is shown in blue with an average value of 2.1\% (over the whole one hour acquisition). This value is not affected by the fiber phase drifts, and hence it is very stable during all the acquisition. 
The contribution due to the phase modulator, shown in orange in the inset, is affected by the random phase drifts and requires stabilization. Indeed, it is possible to recognize the moments when the locking position was lost and see that they are reflected on the overall QBER, shown in red. The average QBER contribution due to the phase modulation and stabilization is of 2.8\% over the whole 1 hour acquisition.\\
%Teoria skr
The QKD protocol implemented in this work is a four-dimensional path-encoded BB84 scheme. It is realized with weak coherent pulses, and hence it requires the integration of the decoy method to counteract an eventual photon number splitting attack. The decoy technique requires that Alice randomly changes the intensity of the quantum states she sends to Bob, choosing among different possible values.
\begin{figure}[ht!]
\centering
\includegraphics[width=0.47\textwidth]{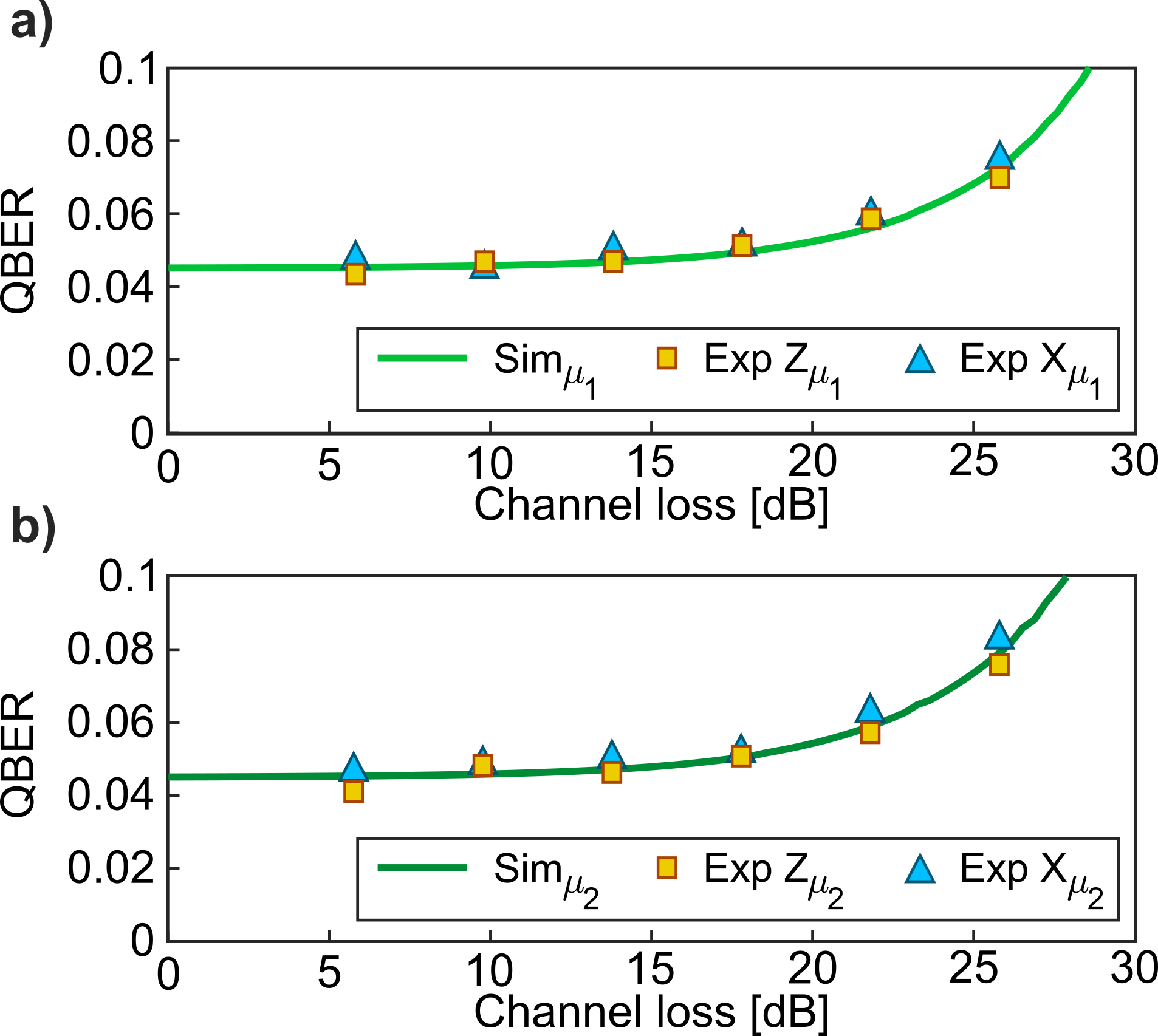}
\caption{Simulation (solid lines) and experimental values for the $\mathcal{Z}$ (yellow squares) and $\mathcal{X}$ (blue triangles) bases for the intensities $\mu_1$, figure a), and $\mu_2$, figure b). Uncertainty values, computed as standard error of the mean, are not displayed as error bars are covered by the markers.}
\label{fig:qberdist}
\end{figure}
For asymptotic key generation regimes, the optimal number of possible intensity values was found to be in general three, but a recent work showed that, when considering finite key regimes (which model more accurately a real implementation), this optimal number is often two~\cite{Rusca2018}.
Moreover, there is a great practical advantage in the implementation of only two levels of intensities in a system: it can be achieved just by adding an IM driven by a squared electrical signal having two possible voltage levels.
\begin{figure}[ht!]
\centering
\includegraphics[width=0.47\textwidth]{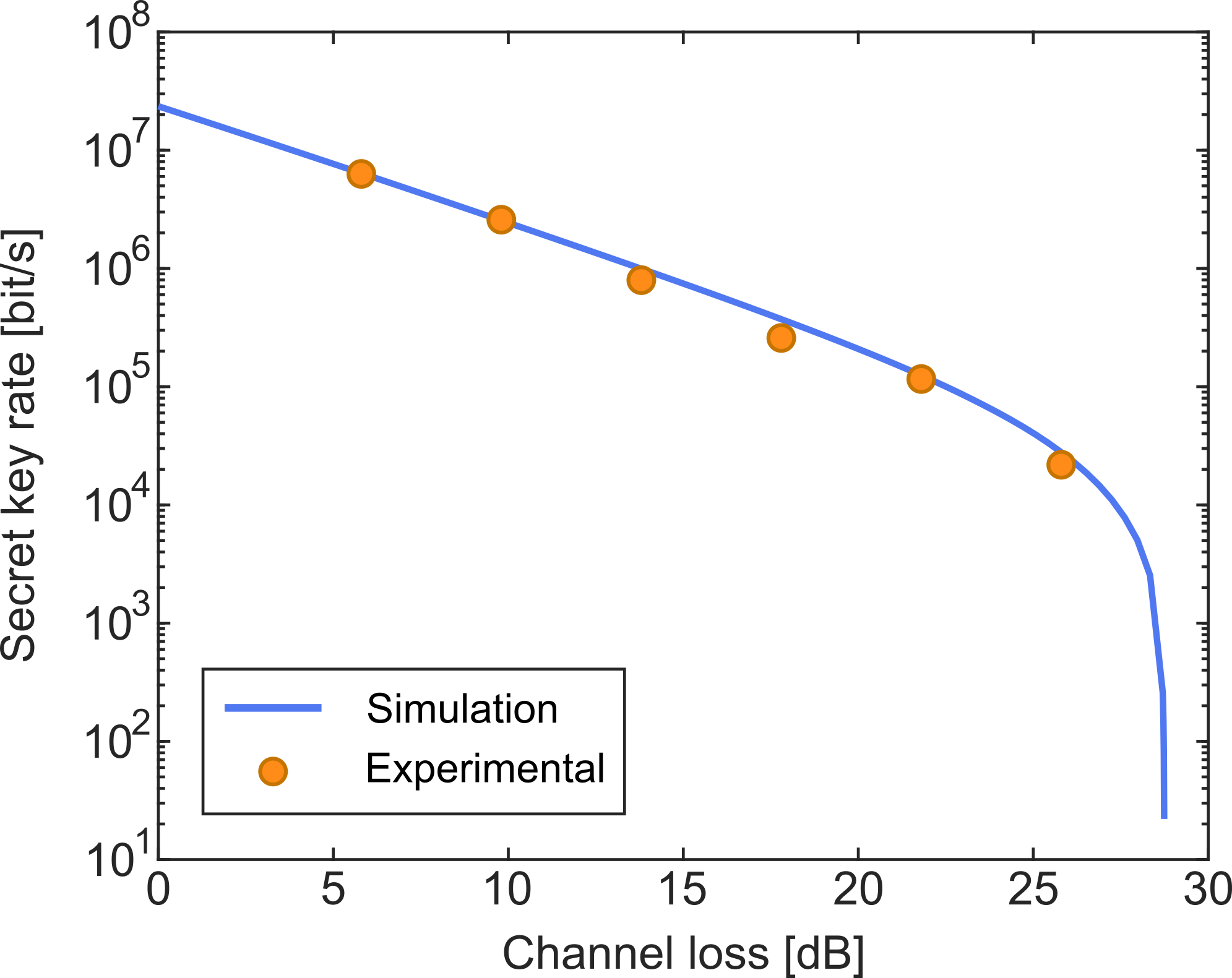}
\caption{Simulation (solid line) of the secret key rate and values obtained from the experimental setup (orange points). Uncertainty values, computed as standard error of the mean, are not displayed as error bars are covered by the markers.}
\label{fig:skr}
\end{figure}
For these reasons, we chose to utilize the secret key bound for a four-dimensional QKD protocol with one decoy (\textit{i.e.} two intensities) in the finite key regime. The equation for the secret key length $\ell$ is~\cite{vagniluca2020}:
\begin{equation}
    \begin{split}
    \ell \ \ \leq \ \  & 2D_0^{\mathcal{Z}} \ \ + \ \ D_1^{\mathcal{Z}} \Bigl[ 2-H\bigl( \phi_{\mathcal{Z}} \bigr) \Bigr] \ \ - \ \ \lambda_{EC} \ \\
    & \ \ - \ \  6\log_2(19/\epsilon_{sec}) \ \ - \ \ \log_2(2/\epsilon_{corr})
    \end{split}
\label{eq:skr}
\end{equation}
The secret key length $\ell$ is defined as the number of secret key bits that are created in a privacy amplification block of length $n_{\mathcal{Z}}$. The terms $D_0^{\mathcal{Z}}$ and $D_1^{\mathcal{Z}}$ are the lower bounds for the vacuum and single-photons events in the $\mathcal{Z}$ basis, respectively; the function $H(\cdot)$ is the high dimensional entropy; the term $\phi_{\mathcal{Z}}$ represents the upper bound on the phase error in the ${\mathcal{Z}}$ basis; $\lambda_{EC}$ is the number of discarded bits during the error correction procedure, and the terms $\epsilon_{sec}$ and $\epsilon_{cor}$ are the secrecy and correctness parameters.
To derive the secret key rate that can be achieved by our setup, we fixed the following values to $n_{\mathcal{Z}}=10^9$ bit and $\epsilon_{sec}=\epsilon_{cor}=10^{-15}$, and we found the optimal values for the two intensity levels $\mu_1$ and $\mu_2$, the probability $p_{\mu_1}$ of sending a state with intensity $\mu_1$, and the probability $p_{\mathcal{Z}}$ with which Alice (Bob) chooses to prepare (measure) in the $\mathcal{Z}$ basis. The values that we used in our implementation are reported in table~\ref{tab:par_result}. With these values, we measured the QBER of the system in the four possible configurations: $\mathcal{Z}$ basis and intensity $\mu_1$, $\mathcal{Z}$ basis and intensity $\mu_2$, $\mathcal{X}$ basis and intensity $\mu_1$ and $\mathcal{X}$ basis and intensity $\mu_2$, registering the number of events as well. These QBER values are reported in table~\ref{tab:par_result}, averaged over 5 minutes of continuous acquisition. With the data collected in such a way, we computed the expected secret key rate using eq.~\eqref{eq:skr}, also shown in table~\ref{tab:par_result}. The secret key rate obtained with our system, after propagation in a 2 km long MCF, is $R_{sk} \approx 6.3$ Mbit/s. This is achievable with a complete system running sessions of approximately 93 seconds long, which is the time to build up the privacy amplification block $n_{\mathcal{Z}}$, a period of time during which the system is efficiently stabilized.
%Simulazione canale maggiore
Furthermore, assuming the signals experience similar phase drifts on longer MCF links, we emulated a longer transmission distance by adding further attenuation to our channel with a VOA. Again, the parameters used for each channel loss, the resulting measured QBER values and the obtained secret key rates are listed in table~\ref{tab:par_result}. As shown in figures~\ref{fig:qberdist} and \ref{fig:skr}, for all four cases the measured QBER and secret key values fit the simulated behaviors.

\section{Discussion}
In this work, we have presented a high-dimensional quantum communication system based on path-encoded states coupled to the cores of a MCF. The faithful transmission of such states is guaranteed by a PLL system, which actively compensates for the random phase drifts acquired during fiber transmission. We implemented a stabilization method that exploits the polarization dependence of the devices used to encode the quantum states. This method allows to maintain a reliable system behavior producing a stable and low QBER for one hour of continuous and free-running acquisition.
Even though temperature changes and polarization drifts affect the long term stability of the system, although the latter happens on a longer time scale than the former, a simple re-calibration of the locking position can recover its optimal performance.
The use of two wavelength multiplexed signals allows for an independent stabilization system that runs simultaneously to the QKD session, but it produces a leakage of 35 kHz from the stabilization to the quantum channel. However, this value does not constitute an issue for low channel losses, as it is orders of magnitude lower than the overall count rate of the quantum states (tens of MHz). Moreover, it is possible to further reduce the leakage by increasing the extinction ratio of the filters. This option usually comes with the price of higher insertion loss, but it might become convenient at higher channel losses.\\
Concerning the QKD protocol realized, a complete implementation requires a real time basis choice and decoy method. However, in our setup both Alice and Bob are only able to prepare and measure the two bases separately. An example of a complete setup would require Alice to use a three optical switches and a different combination of BSs to create all the possible superposition states, while the stabilization system would require the use of 4 PLL boards. The design of an experimental setup allowing for the real-time implementation of our protocol is reported in the Supplementary Information. Finally, to integrate the choice between the two intensity levels for the decoy technique, the addition of a third IM, or of a multi-level signal for the current intensity modulator, is necessary.\\
In addition, we performed a path-encoded 2D-QKD protocol to compare the performance of our high-dimensional scheme with a more standard qubit-based protocol realized with the same setup. The secret key rate achievable is limited to 3.7 Mbit/s, a value lower than the one reached with the 4D scheme. This result demonstrates the actual advantage of using qudit-based protocols over their qubits counterparts. However, as reported in Ref. \cite{bacco2019boosting}, to maximize the achievable secret key rate, a multiplexing approach should also be considered. Indeed, high-dimensional states are beneficial in the case of saturation regime of the single photon detectors, in the case of photon-starved regimes and in situations where the noise in the quantum channel is high enough to prevent the key generation of a qubit scheme \cite{ecker2019overcoming}. Note that this work does not directly falls into these conditions, as demonstrating a high secret key generation is not the main focus of our experiment. Rather, our goal is to show a method that concretely allows for the distribution of high-speed path-encoded states.\\
A further interesting investigation concerns the study of our system using a longer MCF channel. Indeed, we expect that a longer fiber interferometer is affected by faster phase drifts impairing the transmission of the superposition states. Therefore, such study could unveil the potential scalability of our approach.\\
In conclusion, our method largely surpass the limitations of previous experiments based on path-encoded state transmission in terms of channel length, repetition rate and final secret key generation, making our QKD system appealing and comparable in terms of performance with current state of the art systems.

\section{Methods}
\subsection{Experimental setup}
The experimental setup is schematically depicted in figure~\ref{fig:setup} b). The transmitter, Alice, has to prepare the quantum states and send them, together with a stabilization signal, towards the receiver, Bob. Hence, Alice has one continuous wave laser, emitting at 1550.92 nm (coded in red in figure~\ref{fig:setup} b) ), and a second continuous wave laser emitting at 1554.13 nm (coded in blue in figure~\ref{fig:setup} b) ) used for the stabilization channel. The light coming from the first laser, is initially attenuated with a variable optical attenuator (VOA) and then carved into a train of pulses with repetition rate 595 MHz by two cascaded intensity modulators (IMs), only one shown in figure~\ref{fig:setup} b) for clarity, thus creating the required train of weak coherent pulses. Then, Alice prepares all the states belonging to one of the bases in eq.~\eqref{eq_states} with the use of a fast optical switch and two PMLs. For instance, when basis $\mathcal{Z}$ is chosen, the switch either sends a weak coherent pulse to cores 1 and 5 or to cores 2 and 7.
At both switch outputs, a sequence of beam splitter (BS) and PML prepares the superposition among the two chosen cores. Finally, before entering the MCF cores, every path is compensated in length and optical power. The IMs, the switch and the PMs are driven by a field programmable gate array board (FPGA) with a pulsed electrical signal (for the IMs) and squared electrical signals (for the switch and the PMs) which take high or low voltage values based on a pseudo-random binary sequence of seed length 12. The stabilization signal laser output is attenuated by another VOA and sent through all the four cores with three BSs.
Notice that two of the three BSs are shared with the quantum signal: this is crucial for both channels to experience the same phase drifts. Thus, both signals are sent through the transmission channel: the selected cores of a 2 km long 7-core MCF.
The cross-talk is lower than -46 dB between all cores and the measured loss is 5.8 dB in the lossiest core. We take this value as the channel loss and compensate for the difference in the other cores. It must be highlighted that the measured losses largely stem from the fan-in/fan-out devices~\cite{dalio2019}, as MCFs do not exhibit significant loss difference between their cores~\cite{bacco2019boosting,Sasaki2017}, and they are comparable to those of standard SMFs.
Then, the receiver has to measure the quantum states. This is done by projective measurements, \textit{e.g.} if Bob wants to measure in the $\mathcal{Z}$ basis, he puts one BS at the output of cores 1 and 5, and one at the outputs of cores 2 and 7.
At the output of the BSs, Bob separates the two channels using wavelength division multiplexing filters (Fs), directing the stabilization signals to two InGaAs single photon detectors (D5 and D6), one per couple of cores. These are used as reference signals to two PLL boards, each driving a phase shifter (PS) that compensates for detected phase drifts~\cite{dalio2019}. The InGaAs detectors D5 and D6 are set to have an efficiency of 15\% and a dead time of 5 $\mu$s. Indeed, we set the average power of the stabilization channel such that the count rate at the maximum of constructive interference is around 180 kHz, in order to limit saturation effects. The insertion loss on the quantum channel due to the receiver is 2.4 dB, as all elements are fiber based and present very low insertion loss. The superconducting nanowire single photon detectors used for the quantum state measurement (D1 to D4) have 85\% efficiency and 100 Hz dark count rate each. However, the final noise level in the system is increased by the leakage from the stabilization channel: the overall count rate due to the leakage is around 35 kHz.
The counts from the superconductive detectors are collected by a time tagger, which determines the time of arrival of every photon with respect to an electrical synchronization signal coming from the FPGA.\\

\vspace{9pt}

\textit{Note}: During the preparation of this manuscript, the authors became aware of a work by Xiao-Min Hu et al. on a similar topic \cite{hu2020efficient}.

\section*{Data availability}
Data supporting the result presented in the manuscript are available upon request from the corresponding author D. B.

\section*{Code availability}
Codes supporting the result presented in the manuscript are available upon request from the corresponding author D. B.

\section*{Acknowledgements}
\vspace{-0.25cm}
The authors would like to thank D. Rusca for the fruitful discussion. This work is supported by the Center of Excellence, SPOC-Silicon Photonics for Optical Communications (ref DNRF123), by the EraNET Cofund Initiatives QuantERA within the European Union’s Horizon 2020 research and innovation program grant agreement No.731473 (project SQUARE).

\subsection*{Competing interests}
The authors declare that there are no competing interests.

\subsection*{Author contributions}
B. D.~L., D. C. and D. B. proposed the idea.  B. D.~L. and D. C. performed the system experiment. N. B and A. Z. designed and realized the phase-locked loop. B. D.~L. carried out the theoretical analysis on the proposed protocol. D. B. supervised the work. All authors discussed the results and contributed to the writing of the manuscript.

\bibliographystyle{IEEEtran}
%\bibliography{Ref}

%\prtinbibliography

\end{document}